\documentclass[prl,twocolumn,showpacs,amsmath,amssymb,nofootinbib]{revtex4-1}
\usepackage{graphicx}
\usepackage{dcolumn}
\usepackage{bm}
\usepackage{bbm}
\usepackage{ulem}
\usepackage{color}
\usepackage{slashed}
\usepackage{hyperref}
\DeclareMathAlphabet{\mathantt}{OT1}{antt}{li}{it}
\DeclareMathAlphabet{\mathpzc}{OT1}{pzc}{m}{it}

\newcommand{\nc}{\newcommand}

\nc{\be}{\begin{equation}} \nc{\ee}{\end{equation}}
\nc{\bea}{\begin{eqnarray}} \nc{\eea}{\end{eqnarray}}
\nc{\bean}{\begin{eqnarray*}} \nc{\eean}{\end{eqnarray*}}

\begin{document}

\title{Two-terminal transport along a proximity induced superconducting quantum Hall edge}
\author{Oleksandr Gamayun}
\author{Jimmy A. Hutasoit}
\author{Vadim V. Cheianov}
\affiliation{Instituut-Lorentz, Universiteit Leiden, P.O. Box 9506, 2300 RA Leiden, The Netherlands}
\date{\today}

\begin{abstract}
We study electric transport along an integer quantum Hall edge where the proximity effect is induced due to a coupling to a superconductor. Such an edge exhibits two Majorana-Weyl fermions with different group velocities set by the induced superconducting pairing. We show that this structure of the spectrum results in interference fringes that can be observed in both the two-terminal conductance and shot noise. We develop a complete analytical theory of such fringes for an arbitrary smooth profile of the induced pairing. 
\end{abstract}


\maketitle

Superconductivity and the quantum Hall effect are two celebrated phenomena by which quantum physics is manifested at macroscopic scales. Both exhibit universal characteristics insensitive to the microscopic detail. However, the underlying physics are quite different. Superconductivity arises from a spontaneously broken gauge symmetry and is characterized by a local order parameter. In contrast, the quantum Hall effect is attributed to a much subtler non-local topological order. 

Even though each phenomenon has been studied extensively on its own, combining the two in a single hybrid device has been an experimental challenge \cite{Takayanagi1998462,PhysRevB.59.7308,PhysRevB.86.115412,doi:10.1021/nl204415s}. This is because quantum Hall effect requires strong magnetic field, which is abhorred by superconductors. Nevertheless, a stable proximity effect in the quantum Hall regime has been achieved recently \cite{Wan:2015aa,E.:2015aa,Kumaravadivel:2016aa,Amet966,Efetov:2016aa,Lee:2016aa}. The key element of this success is the ability to manufacture a hybrid structure using either superconducting materials with high critical fields or 2-dimensional electron gas that exhibits quantum Hall effects in lower magnetic fields. The approach of Ref. \cite{Wan:2015aa} was to use NbN contacts, with critical fields higher than 16 T, on a 2-dimensional electron gas in a GaAs quantum well. In contrast, in Refs. \cite{E.:2015aa,Kumaravadivel:2016aa,Amet966,Efetov:2016aa,Lee:2016aa}, graphene was used as the 2-dimensional electron gas that exhibits quantum Hall effects in lower magnetic field.
 
\begin{figure}[tb]
\centerline{\includegraphics[width=0.9\linewidth]{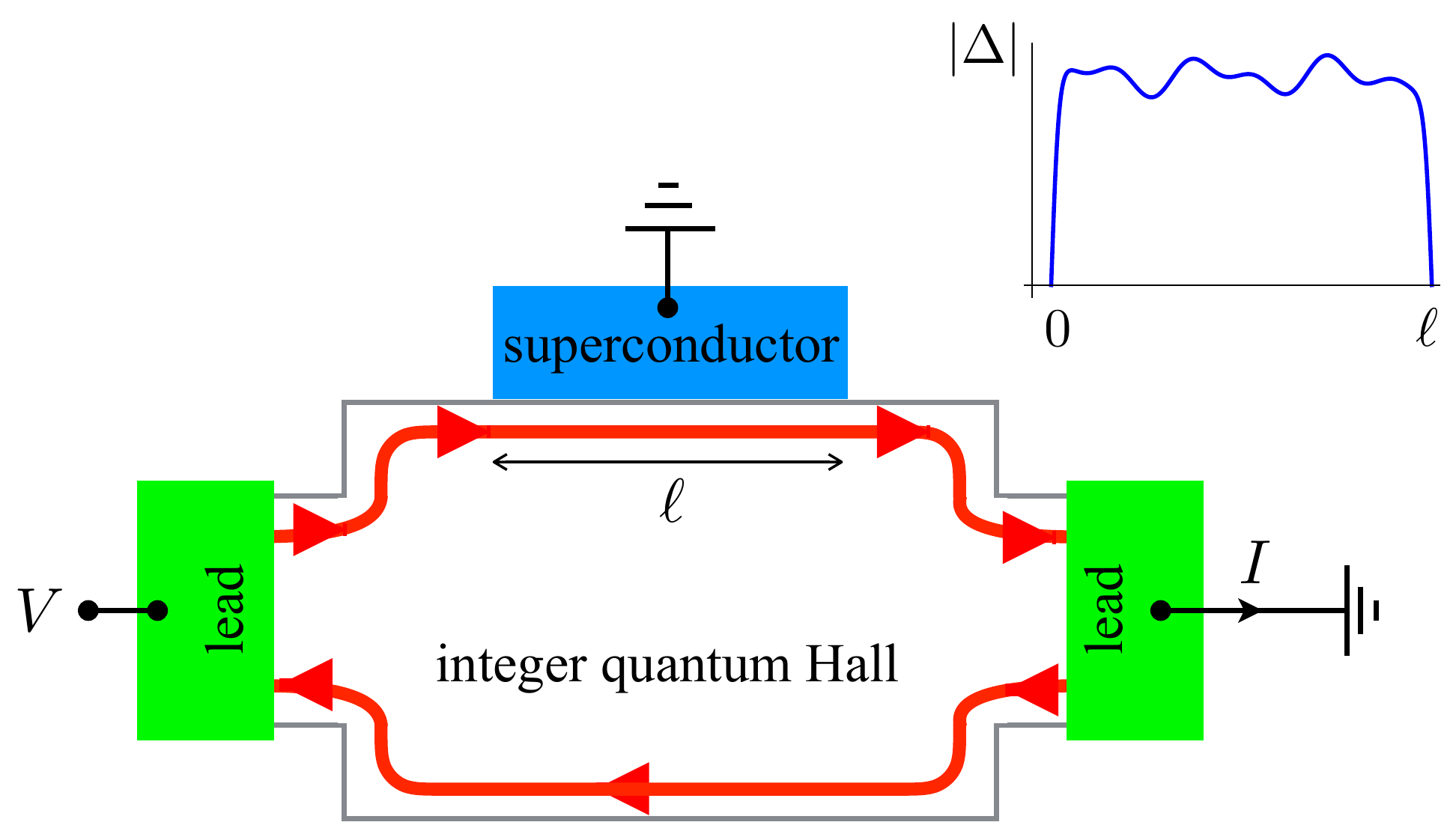}}
\caption{The transport measurement set-up. The quantum Hall--superconductor interface, where the induced pairing is non-vanishing, has a length $\ell$. Voltage $V$ is applied at the upstream lead and the current $I$ is measured at the downstream one. Inset: the induced pairing varies smoothly along the interface but dies rapidly away from the proximity induced superconducting region.}
\label{fig:setup}
\end{figure}

This experimental breakthrough offers an opportunity to test the predictions of earlier theoretical works such as the tunneling current from a superconducting point contact into a quantum Hall liquid \cite{PhysRevB.49.14550} and the critical current \cite{ma1993josephson} along with the upstream information transfer \cite{Huang:2016aa} in a superconductor -- normal metal -- superconductor (SNS) junction, where the normal metal is in the quantum Hall regime. Furthermore, if one can extend the stable proximity effect into the fractional quantum Hall regime, one might be able to create novel excitations with non-trivial braiding statistics \cite{PhysRevX.2.041002,Clarke:2013aa,PhysRevX.4.011036,PhysRevX.5.041042}.

In this letter, we focus on the electric transport along the quantum Hall edge with an extended proximity induced superconducting region. We look into the $\nu =1$ integer quantum Hall case or the situation where only the outermost edge of a $\nu>1$ state contributes to transport.  We consider the geometry\footnote{Ref. \cite{Lee:2016aa} called this ``wide superconductor'' geometry.} depicted in Fig. \ref{fig:setup}, where the induced pairing varies smoothly at the interface but dies rapidly away from the proximity induced superconducting region.  Unlike other proximity induced systems, without fine-tuning, the proximitized edge remains gapless and there are propagating degrees of freedom at zero energy \cite{PhysRevB.83.195441}. When electrons enter the proximity induced region, they are transmitted as two Majorana-Weyl fermions\footnote{As we shall see, strictly speaking, the propagating degrees of freedom in the proximity induced region take the form of relativistic Majorana-Weyl fermions only at high voltage.} with different group velocities. Therefore, it is natural to expect that upon recombination at the end, the current will show an interference pattern akin to the double-slit experiment. This can be thought of as a Mach-Zehnder interference of the co-propagating Majorana-Weyl fermions which uses the intrinsic dynamic of the edge rather than a complicated geometrical set-up in order to create two alternative quantum paths \cite{Ji:2003aa}. It is interesting to draw an analogy between this condensed matter system and the particle physics experiments performed at Super-Kamiokande and Sudbury Neutrino Observatories\footnote{The Nobel prize in physics 2015.}. In the latter, neutrinos are created by weak interaction processes but the propagating degrees of freedom are not the weak interaction eigenstates. The propagating modes have different masses and thus, different group velocities, which results in the interference/oscillation that is measured in the neutrino observatories. In our case, the proximity effect converts the incoming electrons into propagating modes that are not charge eigenstates. Furthermore, the induced pairing behaves like a ``mass" term that even though does not open a gap, results in different velocities of the propagating modes.


Experimentally, the Mach-Zehnder interference will be seen as an oscillatory pattern in the two-terminal conductance and the shot noise. As will be shown below, at large enough applied voltage $V$, both characteristics, as functions of $V$, exhibit oscillations with the period 
\be
\tau = \int\limits \left[\frac{1}{v- \Delta(x)} - \frac{1}{v+ \Delta(x)} \right] dx. \label{eq:scale}
\ee
Here, $v$ is the Fermi velocity, $\Delta$ is the magnitude of the induced pairing and the integral is taken over the proximity induced superconducting part of the quantum Hall edge. The expressions $v \pm \Delta$ can be thought of as {\it local} velocities of the propagating Majorana-Weyl fermions. Thus, $\tau$ is the difference in their times of arrival, which can be determined independently using a time resolved measurement, such as \cite{Glattli:2016aa}. Given the measurement of $\tau$, the two-terminal conductance and shot noise can then be fitted using a single-parameter fit function Eq. \eqref{eq:fit}. At moderate voltage, the picture is more complicated, however, we derive an analytical formula for the current and shot noise at generic $V$, Eq. \eqref{eq:main}. It is worth noting that the voltage $V$ is not the underlying cause behind the interference phenomenon. Instead, it is merely a knob one uses to change the ``length" of the arms in the equivalence Mach-Zehnder interferometer. 

As is shown in \cite{PhysRevB.83.195441}, the most general leading order (in gradient expansion) of the low energy effective Hamiltonian of the proximity induced superconducting quantum Hall edge is given by
\be
H=H_0 + H_{\Delta}, \label{eq:Ham}
\ee
where 
\be
H_0 = -i v \int dx \, \psi^{\dagger}(x)\, \left[\partial_x+ i \partial_x \Omega(x)\right]\psi(x), \label{eq:edge}
\ee
and 
\be
H_{\Delta} = \frac12 \int dx \left[\Delta(x) \psi\left(x\right)\partial_x\psi\left(x\right)+{\rm h.c.}\right],\label{eq:deformation}
\ee
which is a generalization of  \cite{PhysRevB.49.14550}. Here, $\psi$ is a spinless Weyl fermion, $v$ is the Fermi velocity at the quantum Hall edge and $x$ is the natural coordinate along the edge. The magnitude of the induced pairing is given by $\Delta$ and the phase is $2 \Omega$. The first term describes the dynamics of the integer quantum Hall edge, while the second one is the proximity induced pairing\footnote{The term ``induced pairing" is ambiguous and can refer either to the Hamiltonian \eqref{eq:deformation} or to the non-vanishing expectation value $\langle \psi \partial \psi \rangle$ resulting from this Hamiltonian. We would like to clarify that in this article we only use this term in the former sense.}. We note that the Hamiltonian \eqref{eq:Ham} describes a triplet pairing in a spin polarized edge. Therefore, in an experiment using an $s$-wave superconductor, such as \cite{Wan:2015aa,E.:2015aa,Kumaravadivel:2016aa,Amet966,Efetov:2016aa,Lee:2016aa}, spin-orbit interaction is necessary. We expect the magnitude of the induced pairing $\Delta$ to decay rapidly away from the interface region and to vary slowly along the interface, as sketched in Fig. \ref{fig:setup}. 

In order to describe the two-terminal transport, we incorporate the applied voltage as the chemical potential difference between the source and the drain. Furthermore, we neglect the electric field in the proximity induced superconducting region. This is due to the screening by the induced superconductivity as the superconductor sources charges in the form of Cooper pairs. Had there been a penetrating electric field in the proximity induced region, we would have to include a term proportional to $\psi^{\dagger}\psi(x)$ in the Hamiltonian as well.

In existing experimental systems, we expect $\Delta \ll v$ and one might be tempted to think that the effect of $\Delta$ in the two-terminal transport is perturbatively small\footnote{If one were able to tune the ratio $\Delta/v$ to the unity, one would achieve a condition under which one of the Majoranas exhibits a flat band. At this point, thermodynamic observables diverge. That should result, for example, in a sharp peak at the heat capacity as $\Delta$ approaches $v$.}. However, for a long enough interface, the effect turns out to be both non-perturbative and significant. To demonstrate the underlying physics, we first consider a real and homogeneous pairing, which is a good approximation assuming that $v \partial_x \Omega$ is negligible compared to the energy. In this case, we can diagonalize the Hamiltonian in the Bogoliubov basis
\be
c_k= \frac{1}{\sqrt{2}} \left(\psi_k - i \psi^{\dagger}_{-k} \right) \quad {\rm and} \quad
d_k = \frac{1}{\sqrt{2}} \left(\psi_k + i \psi^{\dagger}_{-k} \right), \label{eq:Maj}
\ee
where $\psi_k$'s are the Fourier modes of the electron field with the non-vanishing commutation relation given by $\{\psi^{\dagger}_k,\psi_p\} = 2\pi \delta(k-p)$. In this basis, the Hamiltonian 
\be
H = \int\limits_0^{\infty} \frac{dk}{2\pi}\Big[ \left(v- \Delta\right)k\, c^{\dagger}_k c_k + \left(v+\Delta\right) k \,d^{\dagger}_k d_k \Big] \label{eq:Hamfull}
\ee
is that of two Majorana-Weyl fermions $c_k$ and $d_k$ having the same chirality and traveling at two different velocities $v \pm \Delta$. 

An electron injected into the upstream of the proximitized region will then split into these two Majorana-Weyl fermions. After having traversed the proximitized region, the Majorana-Weyl femions acquire a phase difference $\delta \Phi = V\tau$ where
\be
\tau = \ell \left(\frac{1}{v-\Delta} - \frac{1}{v+\Delta}\right)  \label{eq:simpletime}
\ee
is the time difference between their arrivals at the end of the interface. We note that the voltage $V$ must be smaller than the bulk gaps of both the superconductor and the quantum Hall system.

When $\delta \Phi \approx 2 \ell V \Delta/v^2 > 2 \pi$, one should expect a non-perturbative effect in the form of interference fringes. This expectation is confirmed by a straightforward calculation of the conductance and shot noise, which are given by 
\bea
\frac{dI}{dV} =  \cos \tau V \quad {\rm and} \quad \frac{dP}{dV} =  \sin^2 \tau V, \label{eq:simpleip}
\eea
where $P$ is the noise power as defined in e.g., \cite{nazarov2012quantum}. Here, we use a system of units where $e=h=1$ such that the conductance quantum is given by unity. We note that the interference fringes develop at sufficiently large voltage. Furthermore, the linear response quantities are not affected as can be seen from $dI/dV \to 1$ as $V \to 0$. 

From Eq. \eqref{eq:simpleip}, we see that for high enough $V$, the current becomes negative. This can be understood as follows. The Majorana-Weyl fermions are not charge eigenstates. They are equal superpositions of electrons and holes. Thus, depending on the time difference between their arrivals at the end of the interface, these Majorana-Weyl fermions might recombine into a state that is hole-like (positively charged) rather than electron-like. This results in a negative current. Alternatively, one can understand this via multiple Andreev processes along the interface such that the result at the end of the interface is a hole.

Next, we turn to the more realistic case where the phase and the inhomogeneity of the induced pairing are taken into account. Since the system remains chiral, the steady state current can be calculated by mode matching in the equation of motion for the field $\psi$. As shown in the Supplemental Material, using a certain parametrization, the transport properties for $V>0$ are given by
\bea
\frac{dI}{dV} =\sin 2 \theta \cos \phi \Bigg|_{x=\ell}, \label{eq:cond}
\eea
and
\bea
\frac{dP}{dV} &=& 1-\sin^2 2 \theta \cos^2 \phi \Bigg|_{x=\ell}, \label{eq:shot}
\eea
where $\phi$ and $\theta$ satisfy the ordinary differential equations
\bea
\frac{d\phi}{dy} &=& \frac{V}{V_0} + 2 \cot 2 \theta \, \cos \phi, \nonumber \\
\frac{d \theta}{dy} &=& \sin \phi, \label{eq:bloch}
\eea
with initial conditions $\theta(y=0) = \pi/4$ and $\phi(y=0) = 0$. Here, 
\be
V_0 = v\frac{\sqrt{v^2-\Delta^2}}{\Delta} \frac{d\Omega}{dx}, \label{eq:offset}
\ee
and we have introduced a new coordinate $y\equiv y(x)$ defined as 
\bea
dy = \alpha(x) \, dx, \quad {\rm where} \quad \alpha = \frac{v}{\sqrt{v^2-\Delta^2}} \frac{d\Omega}{dx}.
\eea

We note that $dP/dV + \left(dI/dV\right)^2 = 1$ is a manifestation of the fact that the system remains chiral \cite{sm} and unitary \cite{PhysRevB.53.16390}. Departure from this relationship in an experiment could signal an edge reconstruction, leakage into normal conducting channels or significant inelastic processes.

It is insightful to think of the coupled differential equations \eqref{eq:bloch} as a canonical system with the Poisson bracket
\be
\left\{\phi,\theta \right\} = \frac{1}{\sin 2 \theta},
\ee
and the Hamiltonian
\be
h = \sin 2\theta \cos \phi - \frac{V}{V_0} \cos 2\theta.
\ee
When $\Delta(0<x<\ell)$ and $d\Omega/dx$ are constants, the Hamiltonian $h$ is $y$-independent and the system is completely integrable. Since $h$ is an integral of motion, we can determine its value from the initial conditions and then evaluate $\sin 2 \theta \cos \phi$ at $x = \ell$. This yields
\bea
\frac{dI}{dV} = \frac{V_0^2 + V^2 \cos \tau \sqrt{V_0^2+V^2}}{V_0^2+V^2}, \label{eq:fit}
\eea
where $\tau$ and $V_0$ are defined in Eqs. \eqref{eq:simpletime} and \eqref{eq:offset}, respectively.

\begin{figure}[h]
\centerline{\includegraphics[width=0.9\linewidth]{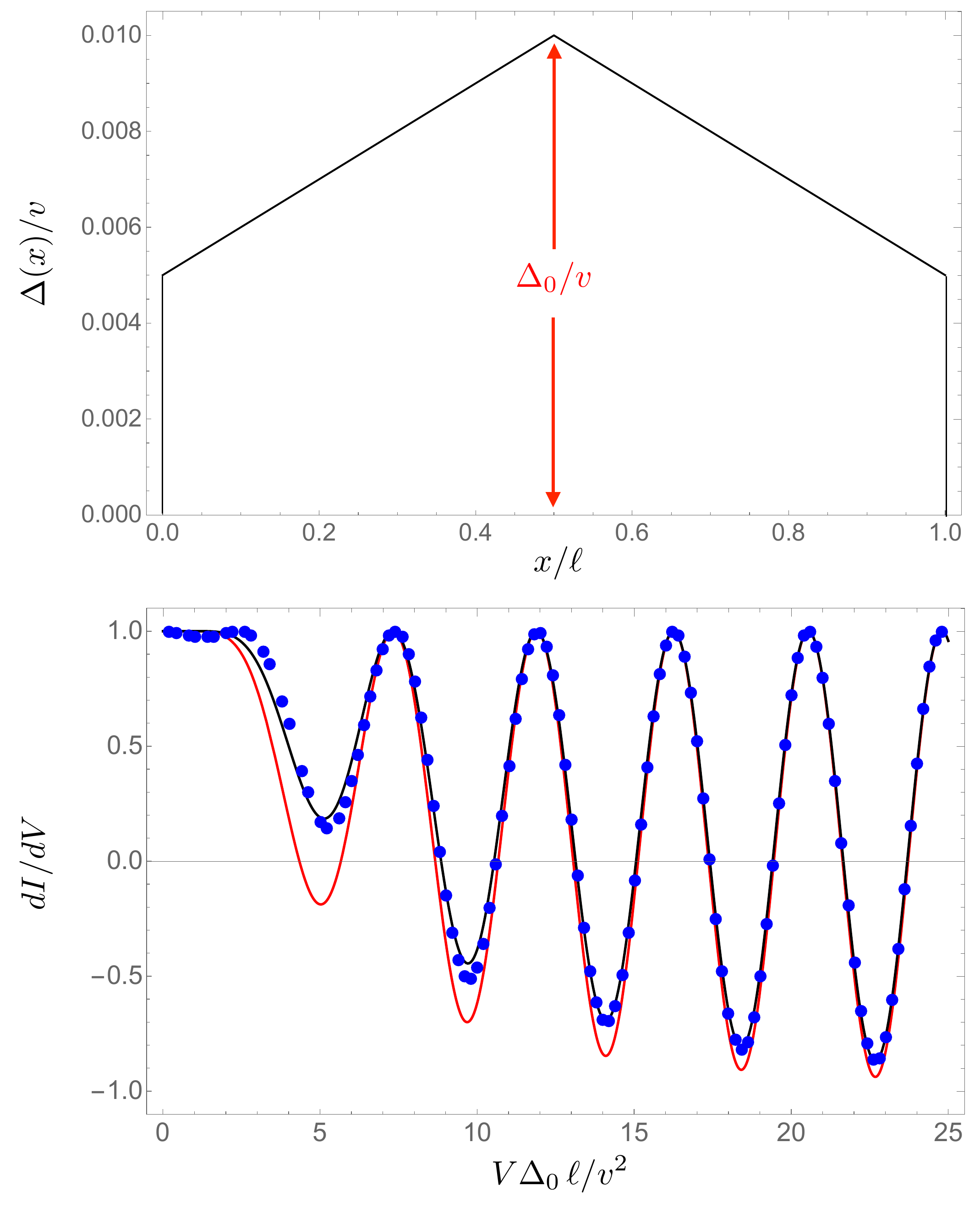}}
\caption{The result for conductance $dI/dV$ as a function of the dimensionless quantity $V \Delta_0 \ell/v^2$ (bottom) for the profile $\Delta(x) = \Delta_0 (1-|1/2-x/\ell|)$ as depicted on the top figure and $\Omega = 3 x/\ell$. We note that the cusp in the figure does not violate the adiabaticity condition \eqref{eq:adcond}. In the bottom figure, the blue dots represent the numerical result while the black line is given by the analytical solution using adiabatic invariant. The red line is the best fitting result using Eq. \eqref{eq:fit}. For the typical Fermi velocity of $10^5$ m/s and the typical length of the proximity induced superconducting region of 10 $\mu$m, one can observe a few periods of oscillation in the conductance by applying a voltage of the order of 1 mV.}
\label{fig:result}
\end{figure}

When $\Delta$ and $d\Omega/dx$ are not constant but slowly varying functions of $x$, we can introduce the adiabatic invariant \cite{landau1976mechanics}, which is given by the integral of the symplectic form
\bea
\mathpzc{J}  = \frac{1}{2\pi}\oint \left(-\frac{\cos 2 \theta}{2}\right) \, d\phi  = \frac{h V_0}{\sqrt{V_0^2 + V^2}}, \label{eq:actionvar}
\eea
i.e., the action variable. The adiabatic invariant is known to be constant for slowly varying parameters. Its canonical conjugate variable, i.e., the angle variable, is given by
\be
\tan \gamma = \frac{V_0 \cos 2 \theta +V \cos \phi \sin2\theta}{ \sqrt{V_0^2 + V^2} \, \sin \phi \sin 2 \theta}, \label{eq:anglevar}
\ee
and it is straightforward to check that indeed $\left\{\mathpzc{J},\gamma\right\} = 1$. The equations of motion and their solutions are then given by
\bea
\frac{d\mathpzc{J}}{dy} &=& \left\{\mathpzc{J},h\right\} \Rightarrow \frac{d\mathpzc{J}}{dx} = 0 \Rightarrow \mathpzc{J} = \frac{V_0(0^+)}{\sqrt{V_0^2(0^+) + V^2}},  \nonumber \\
\frac{d \gamma}{dy} &=& \left\{\gamma, h\right\} \Rightarrow \gamma = \frac\pi 2- \int \limits_0^x \alpha(x') \sqrt{1+\frac{V^2}{V_0^2(x')}} \, dx', \nonumber\\
\eea
which is valid for as long as the following adiabaticity condition holds
\be
\frac{d^2\gamma}{dx^2} \ll \left(\frac{d\gamma}{dx}\right)^2. \label{eq:adcond}
\ee

Using Eqs. \eqref{eq:actionvar} and \eqref{eq:anglevar}, we evaluate $\sin 2 \theta \cos \phi$ at $x = \ell$ and obtain
\begin{widetext}
\bea
\frac{dI}{dV}= \frac{1}{\sqrt{V_0^2(0^+)+V^2} \sqrt{V_0^2(\ell^-) + V^2}} \left\{V_0(0^+) V_0(\ell^-) + V^2 \cos \left[\int \limits_0^{\ell} dx \left(\frac{1}{v-\Delta(x)}-\frac{1}{v+\Delta(x)}\right) \sqrt{V_0^2(x) + V^2}\right]\right\}, \label{eq:main} \nonumber \\
\eea
\end{widetext}
where $V_0(x)$ is defined in Eq. \eqref{eq:offset}. At small $V$, the conductance goes as $dI/dV = 1+ {\cal O}(V^2)$ and at large $V$, it is an oscillating function of $V$ with the period \eqref{eq:scale}. 

A comparison between our analytical solution \eqref{eq:main} and the full numerical solution for a given inhomogeneous profile is shown in Fig. \ref{fig:result}. One can see a remarkable agreement between the two. Yet another remarkable thing is that one can fit this result by using a two-parameter fit function as given in Eq. \eqref{eq:fit}. Furthermore, one can also perform a time-resolved measurement to determine $\tau$ and use Eq. \eqref{eq:fit} as a one-parameter fit function. The fitting result agrees with the exact solution at small and large $V$ while deviates from it at the intermediate values of $V$, see Fig. \ref{fig:result}. 

In regards to the shot noise, one can obtain $dP/dV$ using the formula $dP/dV = 1 - (dI/dV)^2$. It is also an oscillating function of $V$ with the period  \eqref{eq:scale} at large $V$ and it scales as $V^2$ at small $V$. 

It is interesting to draw comparison with other systems that exhibit broken time reversal symmetry and charge non-conservation, namely the chiral $p$-wave superconductor and the proximity induced superconducting quantum anomalous Hall edge. In the former case, neither $dI/dV$ nor $dP/dV$ exhibit interference fringes \cite{PhysRevB.92.121406}. This is because in the chiral $p$-wave superconductor, the pairing does not result in edge excitations having different group velocities. In the latter case, oscillatory behavior is predicted \cite{PhysRevB.93.161401}, albeit of a different physical origin than the present work.

It is important to note that the non-trivial result \eqref{eq:main} assumes discontinuities at $x=0$ and $x=\ell$. In practice, this means that the adiabaticity condition \eqref{eq:adcond} is broken at the ends of the proximity induced region. If the adiabaticity condition \eqref{eq:adcond} is valid everywhere, Eq. \eqref{eq:main} yields $dI/dV=1$, independent of $V$. This is not an artifact of the approximation that lead to Eq. \eqref{eq:main} as the numerical solution to Eq. \eqref{eq:bloch} exhibits identical behavior. 

A loss of adiabaticity does not have to occur only at the ends of the proximity induced region (due to the rapid decay of the induced pairing) but it can also happen throughout the interface due to e.g., impurities. In other words, our result is sensitive to the presence of scatterers that source sharply varying electrostatic potential. This case of dirty edge certainly merits further study.

\if
At the point where adiabaticity is lost, the value of the action variable changes, and once adiabaticity is recovered, the constant nature of the action variable resumes. Therefore, we can extend Eq. \eqref{eq:main} to the case of dirty interface
\begin{widetext}
\bea
\frac{dI}{dV}\propto \frac{1}{2\pi \sqrt{V_0^2(x_{\rm last})+V^2} \sqrt{V_0^2(\ell^-) + V^2}} \left\{V_0(x_{\rm last}) V_0(\ell^-) + V^2 \cos \left[\int \limits_0^{\ell} dx \left(\frac{1}{v-\Delta(x)}-\frac{1}{v+\Delta(x)}\right) \sqrt{V_0(x)^2 + V^2}\right]\right\}, \label{eq:main2} \nonumber \\
\eea
\end{widetext}
where $x_{\rm last}$ is the position of the last scatterer. We note that at small $V$, in general $dI/dV \ne 1/(2\pi)$. This is unlike the case of quantum Hall edge (without proximity effect) where impurities do not affect the edge transport. The difference can be understood as impurity induced leakage of charge carriers into the superconductor.
\fi

\if
In order to understand the limitation of our adiabatic approximation, let us evaluate the correction to the angle variable $\delta\gamma$:
\bea
\partial_y \delta \gamma = \cos^2 \gamma \left( \frac{\partial \tan \gamma}{\partial \frac{\omega}{V_0}} \right)\partial_y \frac{\omega}{V_0} = \frac{\mathpzc{J} \, \cos \gamma}{\sqrt{1-\mathpzc{J}^2}}\frac{\partial_y \frac{\omega}{V_0}}{4+ \frac{\omega}{V_0}^2}. \quad
\eea
We note that this diverges when $\mathpzc{J} = 1$ or when $\frac{\omega}{V_0}(0^+) = 0$. This means that in our adiabatic approximation, we cannot put the pairing amplitude $\Delta$ to zero at the ends of the interface in a smooth enough manner. 
\fi

\textit{Summary and discussion} -- In this letter, we have considered the proximity effect on a clean integer quantum Hall edge. Unlike normal metals, the chiral nature of this system precludes the formation of a gap in the single particle spectrum. Instead, it results in two modes with different group velocities set by the strength of the pairing. At large energy, each of these modes can be described by the Majorana-Weyl Hamitonian. This can be understood from the fact that \eqref{eq:deformation} is a marginal deformation that breaks Lorentz invariance.

Having two modes with different group velocities results in an interference pattern that can be observed in two-terminal transport measurements. In particular, we studied the conductance and shot noise in the case of relatively clean interface where the induced pairing decays rapidly away from the proximity induced superconducting region and varies smoothly within. We found an analytical expression for this generic case, see Eq. \eqref{eq:main}. However, the result can be fitted remarkably well by a simpler two-parameter fit function as given in Eq. \eqref{eq:fit}. We note that the strength of the induced pairing $\Delta$ is unknown and the measurement of the period of the oscillation can determine it. 

\if
Lastly, let us briefly discuss the case of the interface between a fractional quantum Hall system and a superconductor. Unlike the case we have considered here, the proximitized fractional quantum Hall edge cannot be described as a free fermion theory. In general, such an edge is a chiral Luttinger liquid with the Hamiltonian
\be
H_0 = v_{ij} \int dx \, \left(\partial_x \varphi_i\right)\left(\partial_x \varphi_j\right),
\ee
and
\be
H_{\Delta} =   \Delta \int dx  :\cos k_i \varphi_i : \, .\label{eq:Hambos}
\ee
Here, $\varphi_i$'s are the bosonized edge modes, $v_{ij}$ is the velocity/interaction matrix and the coefficients $k_i$'s are chosen such that the vertex operator $:\exp k_i \varphi_i:$ has twice the charge of an electron. In such a system, one might have to abandon the hope to solve the system exactly. Therefore, even in our simple system of a proximity induced integer quantum Hall edge, it is instructive to ask the question how one can reproduce the results we have shown here from perturbation theory. The full story of perturbation theory of quantum Hall edges (or chiral conformal field theories in general) turns out to be subtle. To fully understand it, we would also need to consider a different physical set-up and therefore, we will do so in a separate publication \cite{cpt}.
\fi

\textit{Acknowledgements} -- We thank P. Kim for the discussion on his experimental results. This research was supported by the Foundation for Fundamental Research on Matter (FOM) and the Netherlands Organization for Scientific Research (NWO/OCW)  through the Delta ITP Consortium and by an ERC Synergy Grant.

\setcounter{figure}{0}
\setcounter{equation}{0}
\renewcommand\thefigure{S\arabic{figure}}
\renewcommand\thetable{S\arabic{table}}
\renewcommand\theequation{S\arabic{equation}}

\begin{widetext}
\vspace{1cm}
\begin{center}{\bf Supplemental Material}\end{center}
\hspace{1cm}

In this Supplemental Material, we detail the derivation of Eqs. (9), (10) and (11) of the main text. We base our analysis on the equation of motion 
\bea
\left[\partial_t + v \,\partial_x + i v \,\partial_x\Omega(x) \right] \psi = -  i \Delta \partial_x\psi^{\dagger} -\frac i2 \left(\partial_x\Delta\right) \psi^{\dagger}, 
\eea
which follows from the Hamiltonian (2) of the main text. In the absence of translational invariance, we express the field as
\bea
\psi(x,t) &=& \int\limits_0^{\infty} \frac{d\omega}{2\pi} \, \Bigg\{e^{- i \omega t} \Big[a_{\omega}\,f_{\omega}(x)+b_{\omega}\,u_{\omega}(x)\Big] +e^{ i \omega t} \Big[a^{\dagger}_{\omega}\,g^{*}_{\omega}(x)+b^{\dagger}_{\omega}\,v^*_{\omega}(x)\Big]\Bigg\}, \quad
\eea
where
\be
\left\{a_{\omega}, a_{\omega'}^{\dagger}\right\} = \left\{b_{\omega}, b_{\omega'}^{\dagger}\right\} = \frac{2 \pi}{v} \delta(\omega-\omega'),
\ee
with the boundary conditions $f_{\omega}(x=0) = v_{\omega}(x=0) = 1$ and $g_{\omega}(x=0) = u_{\omega}(x=0) = 0$. Here, $a_{\omega}$ and $b_{\omega}$ can be thought of as electron and hole degrees of freedom.

We will express the solution in terms of 
\be
\chi_{\pm}(x) =  \sqrt{\frac{v}{2\left[v\pm \Delta(x)\right]}} \quad {\rm and} \quad \tau_{\pm}(x) = \int\limits_0^x \frac{dx'}{v\pm \Delta(x')},
\ee
where in the case of vanishing denominator, the integral is understood as a principle value integral. The solution is then given by
\be
   \begin{pmatrix} 
      f_{\omega} e^{- i \omega \tau_-}\\
      i u_{\omega} e^{- i \omega \tau_+} \\
   \end{pmatrix} =    \begin{pmatrix} 
         e^{i\phi_{\omega}}  \cos \theta_{\omega}& \sin \theta_{\omega} \\
         -\sin \theta_{\omega} & e^{-i\phi_{\omega}}  \cos \theta_{\omega} \\
      \end{pmatrix}    \begin{pmatrix} 
            \chi_+ \\
            \chi_- \\
         \end{pmatrix},
\ee
and
\be
   \begin{pmatrix} 
      i g_{\omega} e^{- i \omega \tau_-}\\
      v_{\omega} e^{- i \omega \tau_+} \\
   \end{pmatrix} =    \begin{pmatrix} 
         e^{i\phi_{\omega}}  \cos \theta_{\omega}& -\sin \theta_{\omega} \\
         \sin \theta_{\omega} & e^{-i\phi_{\omega}}  \cos \theta_{\omega} \\
      \end{pmatrix}    \begin{pmatrix} 
            \chi_+ \\
            \chi_- \\
         \end{pmatrix},
\ee
where $\phi_{\omega}$ and $\theta_{\omega}$ satisfy
\bea
\sqrt{v^2-\Delta^2} \partial_x\phi_{\omega} &=& \omega \Delta+ 2 v \partial_x \Omega \cot 2 \theta_{\omega} \, \cos \phi_{\omega}, \nonumber \\
\sqrt{v^2-\Delta^2} \partial_x \theta_{\omega} &=& v \partial_x \Omega \sin \phi_{\omega},
\eea
subject to boundary conditions $\theta_{\omega}(x=0) = \pi/4$ and $\phi_{\omega}(x=0) = 0$. We note that the full solution obeys the correct fermionic (equal time) anticommutation relations.

For steady state, at the large time limit \cite{PhysRevB.53.16390}, we have
\bea
\frac{dI}{dV} &=& |f_V|^2 - |g_V|^2 \bigg|_{x= \ell} = \sin 2 \theta_V \cos \phi_V, \nonumber \\ 
\frac{dP}{dV} &=& \left(|f_V|^2+ |g_V|^2\right)- \left(|f_V|^2- |g_V|^2\right)^2 \bigg|_{x=\ell} = 1 - \sin^2 2 \theta_V \cos^2 \phi_V,
\eea
where our normalization is such that the original Hall conductance (for vanishing induced pairing) is given by unity (in the unit of $e^2/h$). In the last line, we have used $|f_V|^2+ |g_V|^2=1$  which comes from the fact that the system remains chiral and unitary.

\end{widetext}
\bibliography{References}

\end{document}